\documentclass[preprint2,twoside]{hwo}

\usepackage{graphicx}

\input{hwo.h}

\begin{document}

\title{\textbf{\LARGE Probing the variations of interstellar dust abundance and properties within and between galaxies with HWO UV spectroscopy in the Local Volume}}
\author {\textbf{\large Roman-Duval, Julia,$^{1}$ Boquien, Mederic,$^2$ and Choi, Yumi $^{3}$}}
\affil{$^1$\small\it Space Telescope Science Institute, Baltimore, Maryland, USA, USA; \email{duval@stsci.edu}}
\affil{$^2$\small\it Universit\'e C\^ote d'Azur, Observatoire de la C\^ote d'Azur, CNRS, Laboratoire Lagrange, 06000, Nice, France; \email{mederic.boquien@oca.eu}}
\affil{$^3$\small\it NSF National Optical-Infrared Astronomy Research Laboratory, 950 North Cherry Avenue, Tucson, AZ 85719, USA; \email{yumi.choi@noirlab.edu}}


\author{\footnotesize{\bf Endorsed by:}
Narsireddy Anugu (Georgia State University), Michelle Berg (Texas Christian University), Sergei Balashev (Ioffe Institute), Emanuele Bertone (Instituto Nacional de Astr\'ofisica, \'Optica y Electr\'onica, Mexico), Hsiao-Wen Chen (University of Chicago), Christopher Clark (AURA for the European Space Agency/Space Telescope Science Institute), Melanie Crowson (American Public University), Michael Davis (Southwest Research Institute), Annalisa De Cia (European Southern Observatory), Marjorie Decleir (European Space Agency/Space Telescope Science Institute), Vincent Esposito (Chapman University), Gary Ferland (University of Kentucky), Sophia Flury (University of Edinburgh), Luca Fossati (Space Research Institute, Austrian Academy of Sciences), Andrew Fox (AURA for the European Space Agency/Space Telescope Science Institute), Frederic Galliano (CEA Paris-Saclay, France), Ana Ines Gomez De Castro (Universidad Complutense de Madrid), Karl Gordon (Space Telescope Science Institute), Maud Gull (University of California Berkeley), Chamani Gunasekera (Space Telescope Science Institute), Dhanesh Krishnarao (Colorado College), Varsha Kulkarni (University of South Carolina), Eunjeong Lee (EisKosmos (CROASAEN), Inc.), Stephan McCandliss (Johns Hopkins University), Coralie Neiner (Observatoire De Paris), Nikole Nielsen (University of Oaklahoma), Stellar Offner (University of Texas Austin), Patrick Petitjean (Institut d'Astrophysique de Paris), Marc Rafelski (Space Telescope Science Institute), Tanita Ramburuth-Hurt (Wits University), Seth Redfield (Wesleyan University), Adam Ritchey (University of Toledo), Kate Rowlands (AURA for the European Space Agency/Space Telescope Science Institute), Farid Salama (NASA Ames Research Center), Sameer Sameer (University of Oklahoma), Frank Soboczenski (University of York \& King's College London), Grant Tremblay (Center for Astrophysics/Harvard \& Smithsonian), Faustino Vieira (Stockholm University), Daniel Welty (Space Telescope Science Institute)
}



\begin{abstract}
 The cycle of metals between the gas and the dust phases in the neutral interstellar medium (ISM) is an integral part of the baryon cycle in galaxies. The resulting variations in the abundance and properties of interstellar dust have important implications for how accurately we can trace the chemical enrichment of the universe over cosmic time. Multi-object UV spectroscopy with HWO can provide the large samples of abundance and dust depletion measurements needed to understand how the abundance and properties of interstellar dust vary within and between galaxies, thereby observationally addressing important questions about chemical enrichment and galaxy evolution. Medium-resolution (R$\sim$50,000) spectroscopy in the full UV range (950-3150 \AA) toward massive stars in Local Volume galaxies (D $<$ 10 Mpc) will enable gas- and dust-phase abundance measurements of key elements, such as Fe, Si, Mg, S, Zn. These measurements will provide an estimate of how the dust abundance varies with environment, in particular metallicity and gas density. However, measuring the carbon and oxygen contents of dust requires very high resolution (R$>$ 100,000) and high signal-to-noise (S/N $>$ 100) owing to the non-saturated UV transitions for those elements being extremely weak. Since carbon and oxygen in the neutral ISM contribute the largest metal mass reservoir for dust, it is critical that the HWO design include a grating similar to the HST STIS H gratings providing very high resolution, as well as FUV and NUV detectors capable of reaching very high S/N. \\
  \\
  \\
\end{abstract}

\vspace{2cm}

\section{Science Goal}

\indent The science goal of this {\it Habitable Worlds Observatory} (HWO) science case development document (SCDD) is to investigate the variations of the abundance and properties of interstellar dust and metals, which are key  tracers of the interstellar medium and critical components of star formation and galaxy evolution, via HWO ultraviolet spectroscopy.\\

\subsection{Topics Related to the Astro2020:}

This Science Case is relevant to the following Key Science Questions and Discovery Areas of the Astro2020 Decadal Survey Report:

\begin{itemize}

\item {\bf Section F-Q1b} How do Molecular Clouds Form from, and Interact with, Their environment?: ``Over the next decade, we need to develop a comprehensive picture of how dust evolves in the ISM, both in the Milky Way and extragalactic environments. This will require observational constraints as well as theoretical models and simulations that track the dust life cycle. High-resolution UV spectroscopy can provide measurements of the depletion of heavy elements in the MW and nearby galaxies''\\
\item {\bf Section D-Q4c} connecting Local Galaxies to High-Redshift Galaxies: ``Reconciling the metallicity scales of nebular emission lines, neutral gas absorption lines, and stellar photospheric lines will enable the interpretation of the chemical history, transport, and mixing, and the ionization structure of galaxies across cosmic times. These will require tracing the faint UV (HeII, CIII], OIII], SIII], etc.) and optical (HeI, auroral lines) lines within and across HII regions, and measuring abundances and depletion patterns of key elements in the neutral gas and in the photospheres of stars across the full range of metal abundances in nearby galaxies.''\\
\end{itemize}

\subsection{Relevance for Other Broad Scientific Areas in the Astro2020 Decadal Survey:}
\begin{itemize}
    \item Panel on the interstellar medium andstar planet formation
\end{itemize}

\subsection{Rationale}

\indent Over a galaxy's lifetime, metals, the building blocks of solid matter and life, are produced in stars and deposited into the interstellar medium (ISM), the tenuous matter that fills the space between stars in galaxies. These metals cycle between different phases of the ISM: some remain in the gas at different temperatures and pressures, others are locked into dust, and others are ejected into the circumgalactic medium through galactic winds, where they can rain back down into the ISM. This chemical enrichment and incessant cycle of material between stars, interstellar gas and dust, and galaxy halos drives galaxy evolution.\\
\indent Interstellar dust plays a key role in many processes driving galaxy evolution. Dust absorbs stellar light in the optical-ultraviolet and re-radiates this energy in the infrared, thus dominating the radiative transfer of light in galaxies. Dust also shields the interiors of clouds, enabling cooler temperatures favorable to star formation and the formation of molecules. In planet-forming disks around low mass stars, dust grains act as seeds to planet formation. Dust acts as a catalyst for the formation of molecular hydrogen, the dominant component of star-forming molecular clouds. Understanding how the abundance and properties of dust change with environment is therefore crucial to understand star and planet formation, chemical evolution, and galaxy evolution.\\
\indent Metals in the ISM cycle between gaseous and solid dust phases. This cycle results in variations of the abundance and composition of interstellar dust within and between galaxies. How well such variations can be observationally constrained has important implications for galaxy evolution and how accurately we can trace the interstellar medium of galaxies, and track the build-up of dust and metals of the universe and the exchange of metals between stars, gas, and dust over cosmic time. \\
\indent Constraining the abundance and composition of metals and dust in the interstellar medium can only be achieved using ultraviolet (UV) spectroscopy. While the Hubble Space Telescope (HST) has been able to explore the closest galaxies (e.g., the Milky Way and Magellanic Clouds), expanding the exploration of dust properties to the full parameter space and diversity of galaxies will require the sensitivity and multiplexing capabilities of HWO.

\section{Science Objective}

\indent The overall science objective of this HWO SCDD  is to use neutral gas absorption lines and dust depletions to investigate the variations of the abundance and composition of dust and metals as a function of environment. Dust and metals are key tracers of the interstellar medium and a critical component of galaxy evolution through cooling and heating, shielding of star-forming regions, and chemistry.\\
\indent This science case has two specific objectives. The first is to measure neutral gas-phase abundances and dust depletions for constituents of dust and refractory elements (Fe, Si, Mg, Ni, Cr, Mn) and volatiles (Zn, S, P, Kr, Cl, N) in large samples of sightlines toward massive stars in the Local Volume (d$<$10 Mpc). These sightlines will span a broad range of metallicities and gas densities, in order to constrain the variations of the dust abundance and composition within and between galaxies.\\
\indent The second science objective is to measure neutral gas-phase abundances and dust depletions for carbon (C) and oxygen (O), which bear most of the available mass for building dust grains, toward large samples of massive stars in the Milky Way and Magellanic Clouds. \\

\subsection{Rationale}  \label{section_rationale}

\indent  A critical, yet poorly understood component of the baryon cycle in galaxies is the cycling of metals in the ISM between gaseous and solid dust phases, through the depletion of metals from the gas to the dust phase via dust formation, and vice versa, the return of heavy elements from the dust to the gas phase via dust destruction. This cycle results in variations of the abundance and properties of interstellar dust within and between galaxies. The parameters describing the life cycle of metals in the neutral ISM are the dust-to-metal mass ratio (D/M, the mass fraction of metals locked up in dust grains) and the dust-to-gas mass ratio (D/G=D/M x Z, where Z is the metallicity).\\
\indent Because the timescales for dust growth in the ISM are inversely proportional to metallicity and gas density, and because dust is destroyed primarily by shocks, which propagate faster the lower the density of the ISM, density and metallicity are expected to drive variations in the dust abundance and properties. Other parameters, such as radiation field, turbulence, or nearby supernova shocks, may also play a role.\\
\indent Variations in D/M and D/G within and between galaxies, in particular with gas density and total metallicity (gas + dust), and how well such variations can be observationally constrained, have important implications for galaxy evolution and how accurately we can track the chemical enrichment of the universe. Indeed, a comprehensive understanding of the variations of D/M and D/G with local (e.g., gas density, radiation field) and global (e.g., metallicity) environment is required to estimate gas masses based on far-infrared dust emission in both nearby \citep{bolatto2011, schruba2011} and distant \citep{rowlands2014, scoville2023} galaxies. Furthermore, understanding how the fraction of metals locked in dust grains varies with environment (i.e., depletion patterns) is critical to the interpretation of gas-phase abundance measurements in damped Ly$\alpha$ systems (DLAs). DLAs are neutral gas absorption systems with log N(H) $>$ 20.3 cm$^{-2}$ observed over a wide range of redshifts using quasar absorption spectroscopy \citep[e.g.][]{rafelski2012, quiret2016, decia2018}. Thanks to their H I and metallic absorption lines, DLAs trace the chemical enrichment of the universe over cosmic times, and carry the majority of metals at high redshift \citep{peroux2020}. However, gas-phase abundance measurements in DLAs have to be corrected for the depletion of metals from the gas to the dust phase, particularly at metallicities $>$1\% solar. Thus, tracking the chemical enrichment of the universe through DLA spectroscopy requires an understanding of how the fraction of metals in the dust phase varies with metallicity, density, and other environmental factors. These variations can be understood in nearby galaxies, where individual stars and interstellar clouds can be resolved.\\
\indent D/M and D/G can be estimated either using FIR emission to trace dust and HI 21 cm and CO emission to trace gas; or by using absorption line UV spectroscopy of metals, in particular key dust constituents such as C, O, Fe, Si, Mg in neutral gas toward background massive (O and B) stars. FIR-based estimates suffer from large systematic uncertainties on the opacity of dust \citep[e.g.][]{RD2014, RD2017, RD2022a}, motivating the need for estimates based on depletions through UV spectroscopy. Column densities of metals in the neutral interstellar gas can be derived from the absorption line profiles in the UV \citep[between 950 \AA\ and 3150 \AA, see list of transitions in][]{ritchey2023}. The column density of atomic hydrogen can be estimated from the Lyman-$\alpha$ absorption profile, while the column of molecular hydrogen can be derived from modeling the Lyman-Werner bands between 950 \AA\ and 1100 \AA. These combined measurements provide the gas-phase abundances of metals in the ISM. To obtain the dust-phase abundances of those metals, the total (gas + dust) metallicity must be known. This can be achieved through two approaches: Either by assuming that the total elemental abundances equate the photospheric abundances of young stars recently formed out of the ISM \citep[e.g.][]{jenkins2009, RD2021}; or by modeling the abundance patterns of both refractory (e.g., Fe) and volatile (e.g., Zn) elements, as in \citet{ritchey2023, decia2024, konstantopoulou2024}. The depletion $\delta$(X) of an element X is then the logarithm of the fraction of that element in the gas phase. The fraction of that element locked in dust is then (1-10$^{\delta(\rm(X)}$).\\
\indent Currently, HST has obtained relatively large samples of depletions for all elements, except C, in the Milky Way. Large samples for most elements (but not C, O, and other lower abundance elements) have been acquired by HST in the Magellanic Clouds. Only a few sightlines in the closest very low metallicity galaxies (e.g., Sextans A, IC 1613) have been obtained with HST, and the S/N of the spectra only allows for measurements of abundances for S and Fe \citep{hamanowicz2024}.  Photospheric abundances of young stars in the Local Volume (d$<$10 Mpc), which are the proxy for the total ISM metallicity (gas + dust), will be obtained by the forthcoming extremely large telescopes (ELTs). Samples already exist for galaxies within 2 Mpc from the VLT and GMT \citep[e.g.][]{kaufer2004, bresolin2007, evans2007}. \\
\indent HWO multi-object medium-resolution (R$\sim$50,000) UV spectroscopy and single- or multi-object high-resolution (R$>$100,000) spectroscopy will be uniquely capable of obtaining large samples of UV spectra toward massive stars in Local Volume galaxies (d $<$ 10 Mpc). These spectra will enable precise measurements of neutral gas-phase interstellar abundances and depletions for the full suite of dust constituents, and over a wide metallicity and ISM density parameter space, thus providing important constraints on the variations of D/G and D/M in the nearby universe. In addition, high-spectral resolution UV spectroscopy will enable the detailed study of chemical abundances not only along the full line of sight, but also in individual ISM clouds (``absorption line components''), Those will in turn inform our understanding of the complex exchanges of metals that regulate the baryon cycle, through enrichment and mixing. 
\\

\begin{figure*}
\begin{center}
\includegraphics[width=\textwidth]{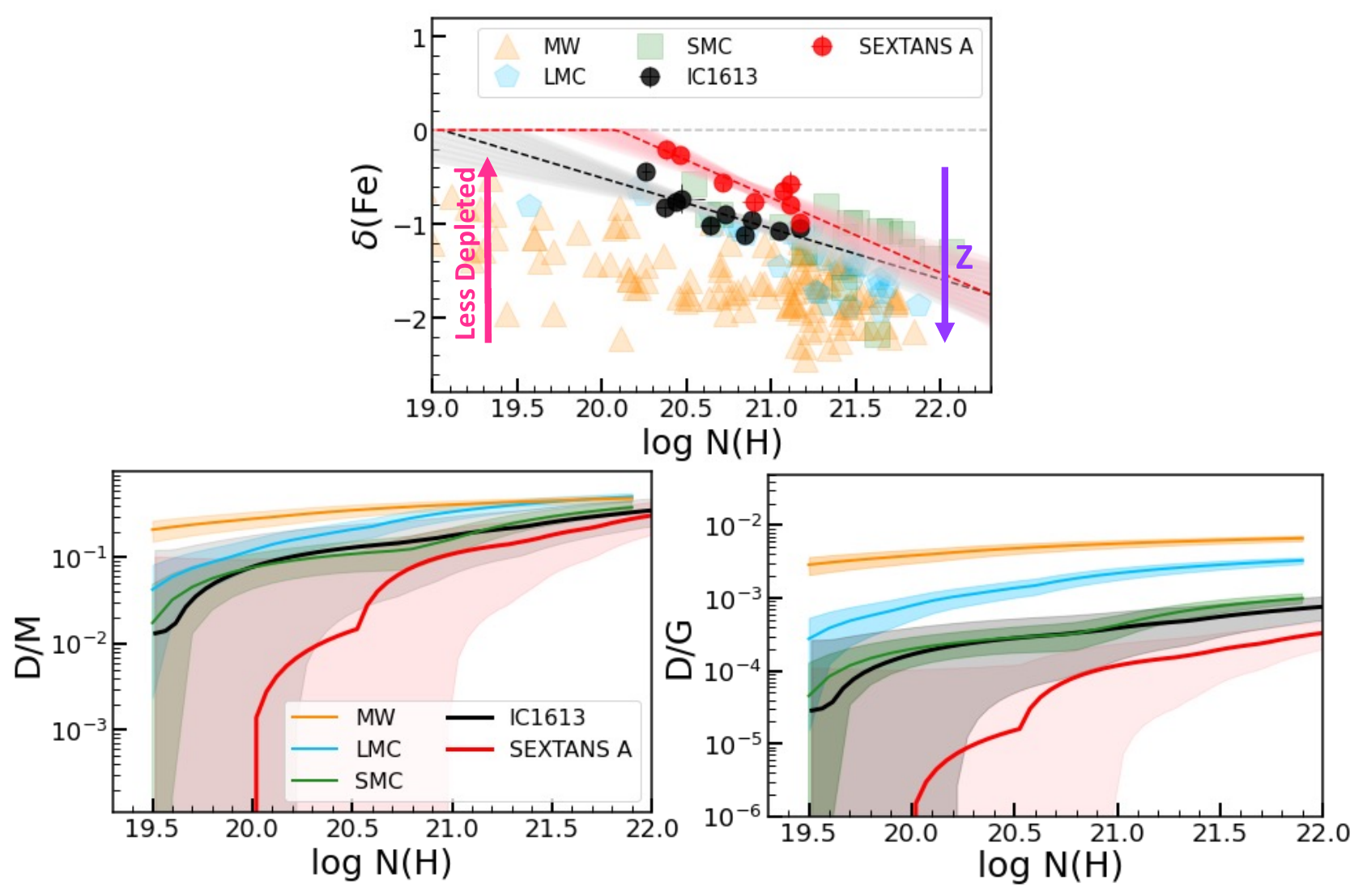}
\caption{\small (from \citet{hamanowicz2024}: (Top) Depletion (=log gas phase fraction) of iron as a function of hydrogen column density in the Milky Way (orange), LMC (blue, 50\% solar), SMC (green, 20\% solar)), IC 1613 (black, 10-20\% solar), and Sextans A (red, 7\% solar). This panel shows that, as the gas density increases, metals become more depleted onto dust, thus increasing the D/G and D/M. (Bottom) Dust-to-metal ratio (left) and dust-to-gas ratio (right) as a function of log N(H) in Sextans A (red), IC 1613 (black), the SMC (green), LMC (blue), and MW (orange). D/M and D/G are derived from depletions obtained with UV spectroscopy toward background O and B stars. In IC 1613 and Sextans A, only Fe depletions can be measured with HST. To extrapolate D/G and D/M from Fe depletions, the relation between the depletions of Fe and other elements established in the Milky Way is assumed, which contributes to the large uncertainty (transparent bands, 1$\sigma$ uncertainty). The metallicity drops from solar at MW on the top to 7\% solar with Sextans A at the bottom. D/M decreases with decreasing density and metallicity. As a result, D/G decreases faster than metallicity.
\label{fig1}
}
\end{center}
\end{figure*}

\subsection{Rationale for science objective \#1}

\indent UV spectroscopy with the COS and STIS instruments on Hubble has enabled robust measurements of depletions in the Milky Way and Magellanic Clouds (20-100\% solar metallicity) for the main components of dust (Mg, Si, Fe, Ni, Cr) and metallicity tracers (S, Zn), see \citet{jenkins2009, tchernyshyov2015, jenkins2017, RD2021, RD2022a}. Measurements obtained with HST probe the column density range log N(H) = 20-22 cm$^{-2}$ in those galaxies with dozens of sightlines. Only with dozens of sight-lines can a statistically robust ($>$5$\sigma$) characterization of the dependence of depletions on environment, specifically gas density and galaxy metallicity, be obtained \citep{RD2021, RD2022a}. Measurements in the Milky Way, LMC, and SMC were therefore obtained mostly through large programs, e.g., GO-14675 and more recently the ULLYSES Director Discretionary Time program, which utilized 450 orbits dedicated to the LMC and SMC. These programs have shown that 1) the depletions of different elements exhibit a high degree of correlation, suggesting a common physical origin for the collective depletion behavior of metals; 2) the level to which an element is depleted in dust increases with increasing gas density, resulting in a factor 5-10 increase in D/G from the diffuse to the dense ISM (log N(H) = 20-22 cm$^{-2}$, Figure \ref{fig1}, bottom right); and 3) that the fraction of metals locked in dust decreases with decreasing metallicity, resulting in the D/M being a factor of 2 lower in the SMC compared to the Milky Way (Figure \ref{fig1}, bottom left). Note that the D/G in the LMC and SMC was based on depletion measurements for key constituents of dust, such as Fe, Si, Mg, but not C and O, which cannot be detected with HST beyond the Milky Way (see objective \#2). To estimate D/G in those galaxies, C and O depletions were estimated by scaling Fe depletions using the relation between the depletions of Fe, C and O established in the Milky Way \citep{jenkins2009}. This results in large statistical errors on D/G \citep[given the large error on the Milky Way relation between Fe, C, and O depletions, see, e.g., Figure 7 in][]{RD2022a}. In addition, there is no reason why the relation between Fe, C and O depletions should hold at low metallicity, where C/O might deviate from the Milky Way value, resulting in different chemical affinities. The assumption of metallicity invariance of the relation between the depletions of different elements may therefore not be valid, resulting in a large systematic uncertainty for C and O depletions, which are only measured in the Milky Way. \\
\indent In addition, depletion corrections based on the [Zn/Fe] abundance ratio have been derived from HST COS and STIS observations in the MW, LMC, and SMC (20-100\% solar metallicity) for all main constituents of dust \citep[except C and O, see][]{decia2016, RD2022b}. These depletion corrections are being used to correct DLA gas-phase metallicities for dust depletion effects out to redshift 5 \citep{RD2022b, welsh2024}.\\
\indent Beyond the Milky Way and Magellanic Clouds, only a few stars and elements with transitions in the FUV (S, Fe) can be reached by the more sensitive COS instrument in lower metallicity galaxies within 2 Mpc. In Sextans A \citep[d = 1.3 Mpc, 7\% solar metallicity][and references therein]{hamanowicz2024}, only 8 stars have been observed in the 1150-1400 \AA\ range with a total of 90 orbits, highlighting the high cost of such observations \citep[][and Figure \ref{fig1}, top]{garcia2014, zheng2020, hamanowicz2024}. The measurements sparsely probe the range log N(H) = 20.3-21.3 cm$^{-2}$. As a result, error bars on the slope of Fe depletions as a function of log N(H) are large \citep[see Table 7 in][]{hamanowicz2024}. The COS/FUV and NUV performance in the 1400-2500 \AA\ range is not sufficient to observe key elements (Si, Fe, Zn, Cr) beyond the SMC. In Sextans A and based on 8 sightlines, the Fe depletions suggest that the trend of D/M increasing with increasing gas density, and decreasing with metallicity continues below the 20\% solar metallicity of the SMC (Figure \ref{fig1}). However, these measurements need to be confirmed for a broader suite of elements and sightlines. Specifically, depletions for key constituents of dust such as Si, Mg, Ni, as well as volatiles that can trace depletions and metallicity, such as S, Zn, and Kr, are required. \\
\indent Within a given galaxy, large samples of sightlines (dozens) are required to sample the full range of neutral gas column density (N(H) = 10$^{20}$-10$^{22}$ cm$^{-2}$) as well as other local conditions (radiation field, dynamic environment, etc.). Only HWO's sensitivity (10x-100x HST) and multi-object medium-resolution (R$\sim$50,000) spectroscopic capability can reach such large samples of sightlines in galaxies with a range of metallicities over the broad UV wavelength range covering transitions of key elements.\\

\subsection{Rationale for science objective \#2}

\begin{figure*}
\begin{center}
\includegraphics[width=0.6\textwidth]{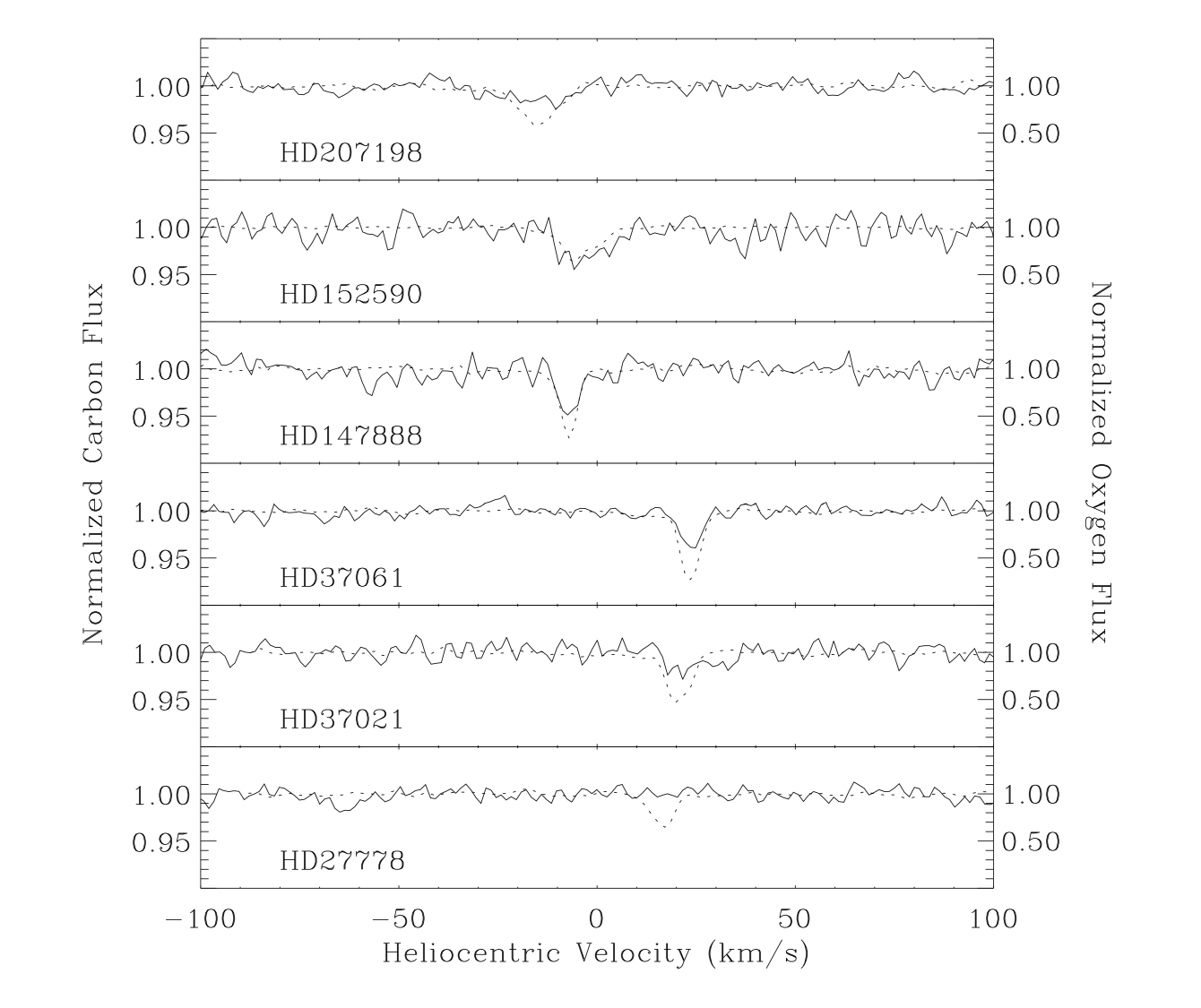}
\caption{\small (from \citet{sofia2004}) Normalized HST STIS echelle spectra of the weak C II] $\lambda$2325 (solid line) and O I $\lambda$1356 (dotted line) absorption features. The normalized flux scale is on the left for carbon and the right for oxygen. Robustly detecting those transitions and measuring O and C abundances requires high-resolution ($>$100,000) and high S/N ($>$100/pixel).
\label{fig2}
}
\end{center}
\end{figure*}

\indent Despite carbon and oxygen being the largest reservoir of mass available for interstellar dust growth, their abundance in neutral gas and dust have only been measured with HST/STIS spectroscopy toward the few brightest Milky Way stars \citep[Figure \ref{fig2}, ][]{sofia2004}. The reason for this is that the C II line at 1335 \AA\ is heavily saturated, while the C II 2325 \AA\ transition is extremely weak, therefore requiring very high S/N ($>$100) and spectral resolution ($>$100,000), as demonstrated in \citet{sofia2004} and references therein (Figure \ref{fig2}). The abundance of oxygen in neutral gas and dust is also difficult to measure because the 1302 \AA\ line is also heavily saturated, while the 1355 \AA\ line is very weak (though not as weak as the C II 2325 \AA\ line). The neutral gas abundance and depletion of O only measured the Milky Way with relatively large error bars, but not in extragalactic environments. The C and O budget of dust and its variations with metallicity are therefore poorly constrained.  \\
\indent In addition, the abundance of Si in dust is constrained only by the single $\lambda$1808 \AA\ transition of Si II, the dominant ion in neutral gas (other Si II transitions at shorter wavelengths are heavily saturated). As a result, unresolved and/or mild saturation cannot be accounted for, resulting in larger uncertainties in Si abundances in neutral gas and dust. Measuring Si abundances more accurately would require detection and line profile analysis of a second Si II transition at $\lambda$2335 \AA\, which is extremely weak, and therefore requires high resolution and high S/N. \\
\indent Given that C, O, and Si bear the majority of the metal mass in dust and available for dust growth. these limitations in measuring C, O, and Si abundances result in a poor understanding of how the composition and abundance of dust changes with environment, in particular density and metallicity. This gap in our understanding of the build-up of dust limits our ability to measure dust and gas masses across cosmic time using far-infrared emission as a tracer of the ISM.\\
\indent High-resolution (R $>$ 100,000), high-S/N ($>$100) UV (1150-3150 \AA) spectroscopy with HWO in the Milky Way and Magellanic Clouds will provide the constraints on the C, O, and Si budget of neutral gas and dust down to metallicity 20\% solar. These constraints will critically improve our general understanding of the cosmic build-up of dust.\\
\indent R$>$100,000 UV spectroscopy with HWO will enable a broader range of impactful studies of chemical abundances in the ISM of galaxies beyond the Milky Way. Recent HST/STIS programs already capitalize on the instrument's highest spectral resolution (R=110,000-140,000) to study ISM cloud-to-cloud variations in chemical properties from individual components of the absorption-line profiles, but this is currently possible only towards the brightest stars in the Milky Way. Such studies are informing chemical enrichment and mixing in the ISM, and enabling a search for the chemical signatures  of very massive stars in interstellar gas.

\subsection{Physical Parameters} \label{sec_phys_param}

\indent Neutral gas-phase abundances of various elements will be derived through analysis of absorption lines for H (Ly$\alpha$), H$_2$ (Lyman-Werner bands), C, N, O, Mg, Si, Si, P, S, Cl, Cr, Mn, Fe, Ni, Cu, Zn, Ge, Kr  \citep[see list of transitions in ][]{ritchey2023}. Large samples of sightlines probing through a broad range of hydrogen column densities in a large sample of galaxies in the Local Volume (d $<$ 10 Mpc) with a broad range of metallicities and gas densities will be targeted. Depletions will be derived from the combination of UV-based neutral gas phase abundances and photospheric abundances in young stars obtained from ground-based spectroscopy with the ELTs.\\
\indent Two techniques for recovering abundances and dust depletions have been applied extensively to HST data (see Section \ref{section_rationale}). The first \citep{jenkins2009, tchernyshyov2015, RD2021, jenkins2017} uses photospheric abundances of young stars as a proxy for the total ISM metallicity to infer the fraction of metals in dust from UV absorption line spectroscopy of neutral gas. The second uses neutral gas abundance patterns for a range of refractory (e.g., Fe, Ni, Cr) and volatile (e.g., O, Zn, S, Kr) elements to derive total abundances and depletions \citep{ritchey2023, decia2024, konstantopoulou2024}.\\
\indent In both cases, large samples UV spectroscopy toward O and early B stars, which have fewer and weaker photospheric lines to disturb the continuum, are needed. Physical parameters for both science objectives are summarized in Table \ref{tab:physical_parameters}. \\

\begin{table*}[!ht]
\begin{center}

    \begin{tabular}{p{0.19 \linewidth} |  p{0.19\linewidth}|  p{0.19\linewidth}  | p{0.19\linewidth}  | p{0.19\linewidth}  }
Physical Parameter & State of the Art  & Incremental Progress (Enhancing) &Substantial progress (enabling)  & Major progress (breakthrough)    \\
\hline
&&&& \\
Number of sightlines with well determined depletions for C and O  &  $\sim$12 (MW)  \newline \newline  0 (other galaxies) & $\sim$50 (MW) \newline \newline  0 (other galaxies) & $\sim$100 (MW) over log N(H) = 20-22 cm$^{-2}$ \newline \newline 10 per galaxy (LMC, SMC)  \newline \newline   0 (other galaxies) & Same as enabling for MW \newline \newline  $\sim$100 over log N(H) = 20-22 cm$^{-2}$ (LMC and SMC) \newline \newline 0 (other galaxies) \\
\hline
&&&& \\
Number of sightlines with well determined depletions for all other main dust constituents (e.g., Fe, Si, Mg, Cr, Ni etc.) &  $\sim$100 (MW)   \newline \newline  $\sim$35 (LMC)  \newline \newline     $\sim$20 (SMC)   \newline \newline Sightlines cover log N(H)  =20-22 cm$^{-2}$   \newline \newline  0 (other galaxies)  \newline \newline    NB: larger samples of depletions are being derived in the LMC/SMC from ULLYSES   &  $\sim$100 (MW)    \newline \newline  $\sim$100 (LMC)   \newline \newline  $\sim$100 (SMC)    \newline \newline  Sightlines cover log N(H)  =20-22 cm$^{-2}$    \newline \newline   0 (other galaxies) &  Same as incremental progress for MW, LMC, SMC   \newline \newline    $\sim$30 per galaxy over log N(H) = 20-22 cm$^{-2}$ out to a few Mpc to probe very low metallicity galaxies ($<$0.1 solar)   & Same as incremental progress for MW, LMC, SMC   \newline \newline     $\sim$100 per galaxy over log N(H) = 20-22 cm$^{-2}$ out to 10 Mpc to probe the full diversity of galaxies (metallicity, morphology, dynamics etc.) \\
\hline
&&&& \\
Number of sightlines with at least Fe depletion measurements & Same as above for MW, LMC, SMC   \newline \newline   $<$10 (IC 1613)  \newline \newline   $<$10 (Sextans A) &  Same as above for MW, LMC, SMC   \newline \newline    $\sim$30 (IC 1613)    \newline \newline   $\sim$30 (Sextans A)   \newline \newline   Over log N((H) = 20-22 cm$^{-2}$ & N/A (incremental progress only)  & N/A (incremental progress only) \\
\end{tabular}
\end{center}
\caption{Physical parameters probed by this SCDD, for state-of-the-art, as well as incremental, substantial, and major progress scenarios.}
\label{tab:physical_parameters}
\end{table*}

\subsection{Physical parameters for science objective \#1}

\begin{figure*}[ht]
\begin{center}
\includegraphics[width=0.8\textwidth]{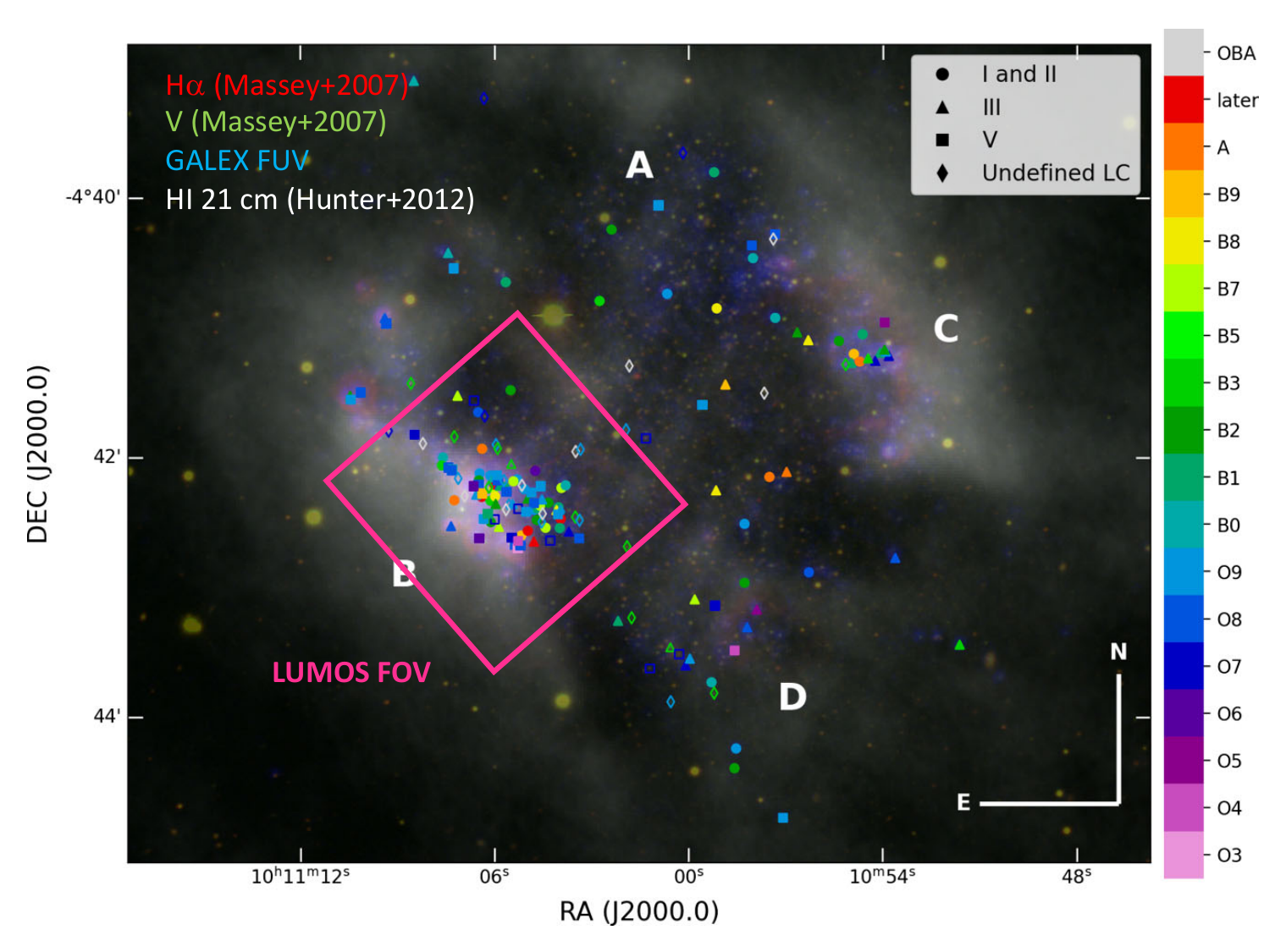}
\caption{\small (From \citet{lorenzo2022}) RGB composite image of Sextans A made with H$\alpha$ (red) and V bands (green) from \citet{massey2007}, and GALEX FUV (blue). The LITTLE THINGS neutral hydrogen map \citep{hunter2012} is overlaid in white. OB stars catalogued in \citet{lorenzo2022} are color-coded according to their spectral type and with different symbols based on their luminosity class. The LUMOS field-of-view is overlaid. In 100h, HWO could obtained FUV and NUV spectra of most of the stars shown in this image with S/N $>$ 20.
\label{fig3}}
\end{center}
\end{figure*}

\begin{figure*}
\begin{center}
\includegraphics[width=8cm]{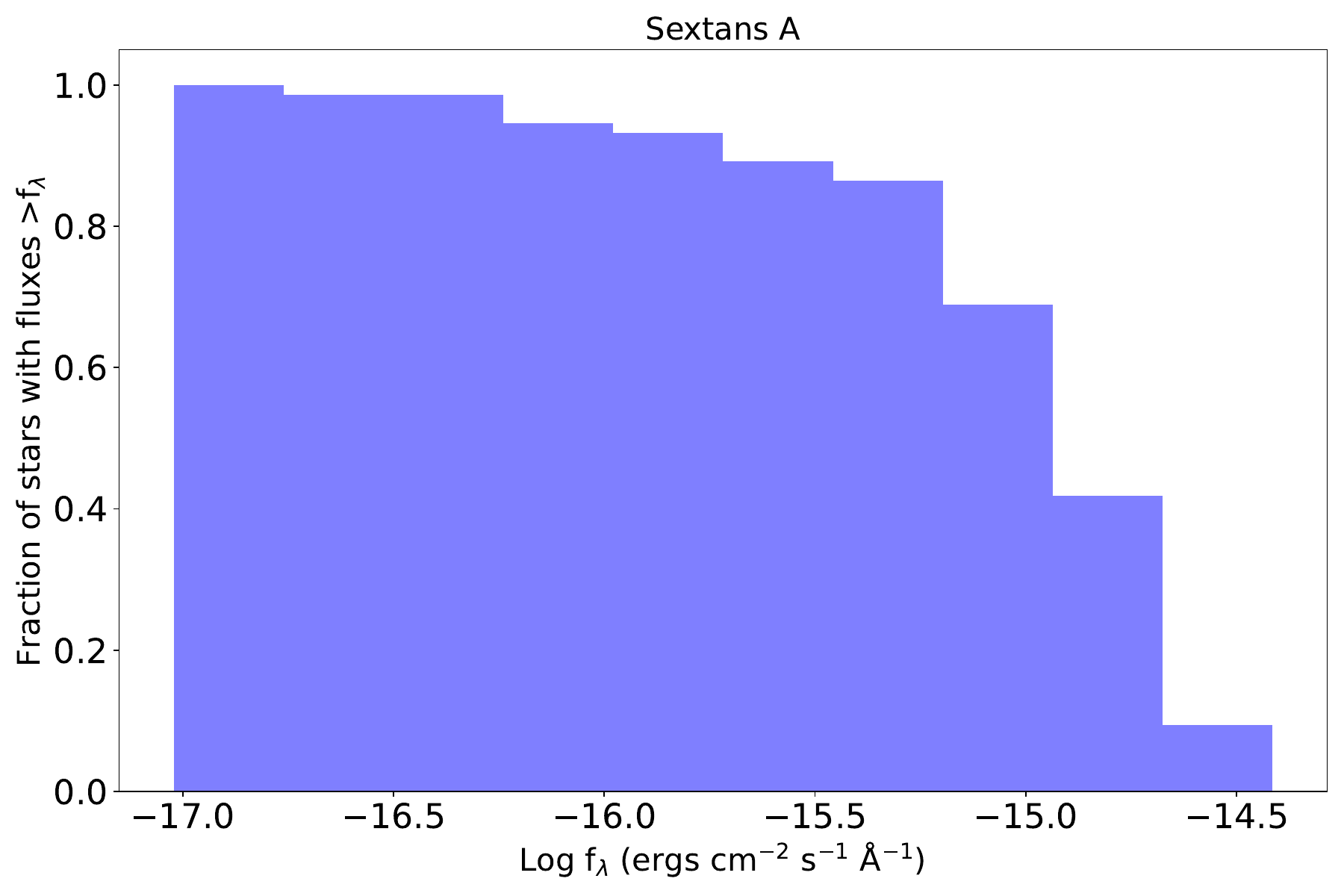}
\includegraphics[width=8cm]{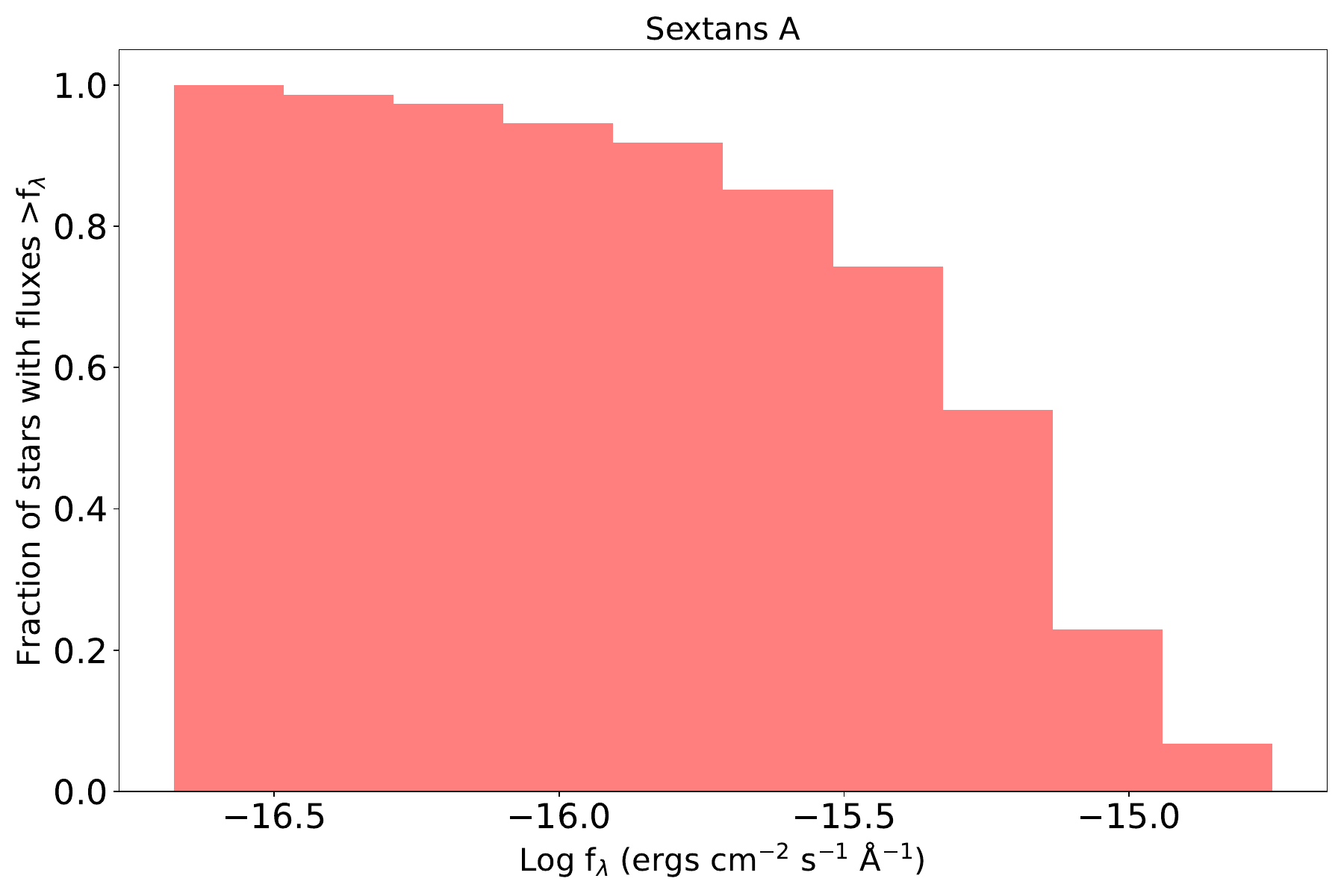}
\includegraphics[width=8cm]{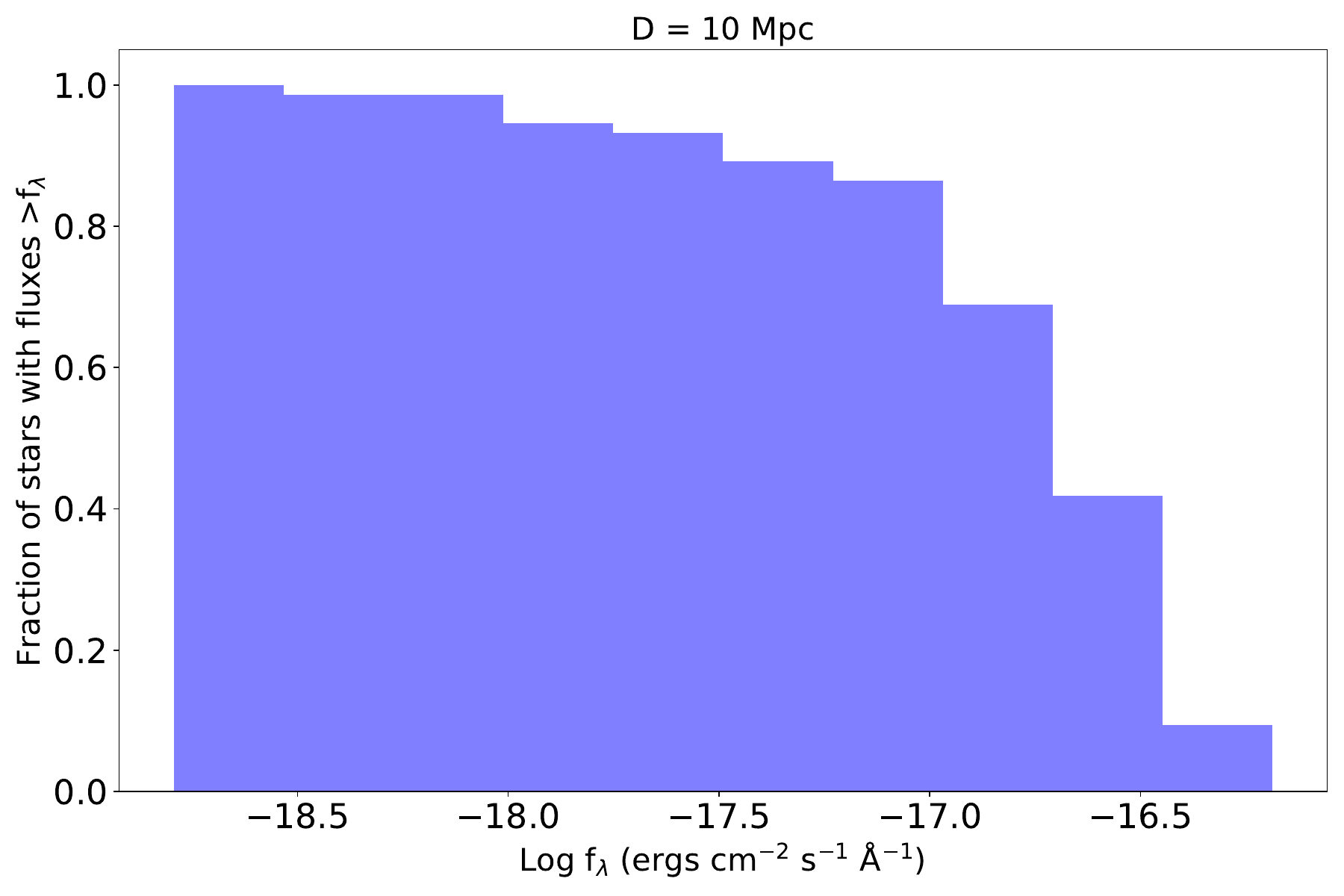}
\includegraphics[width=8cm]{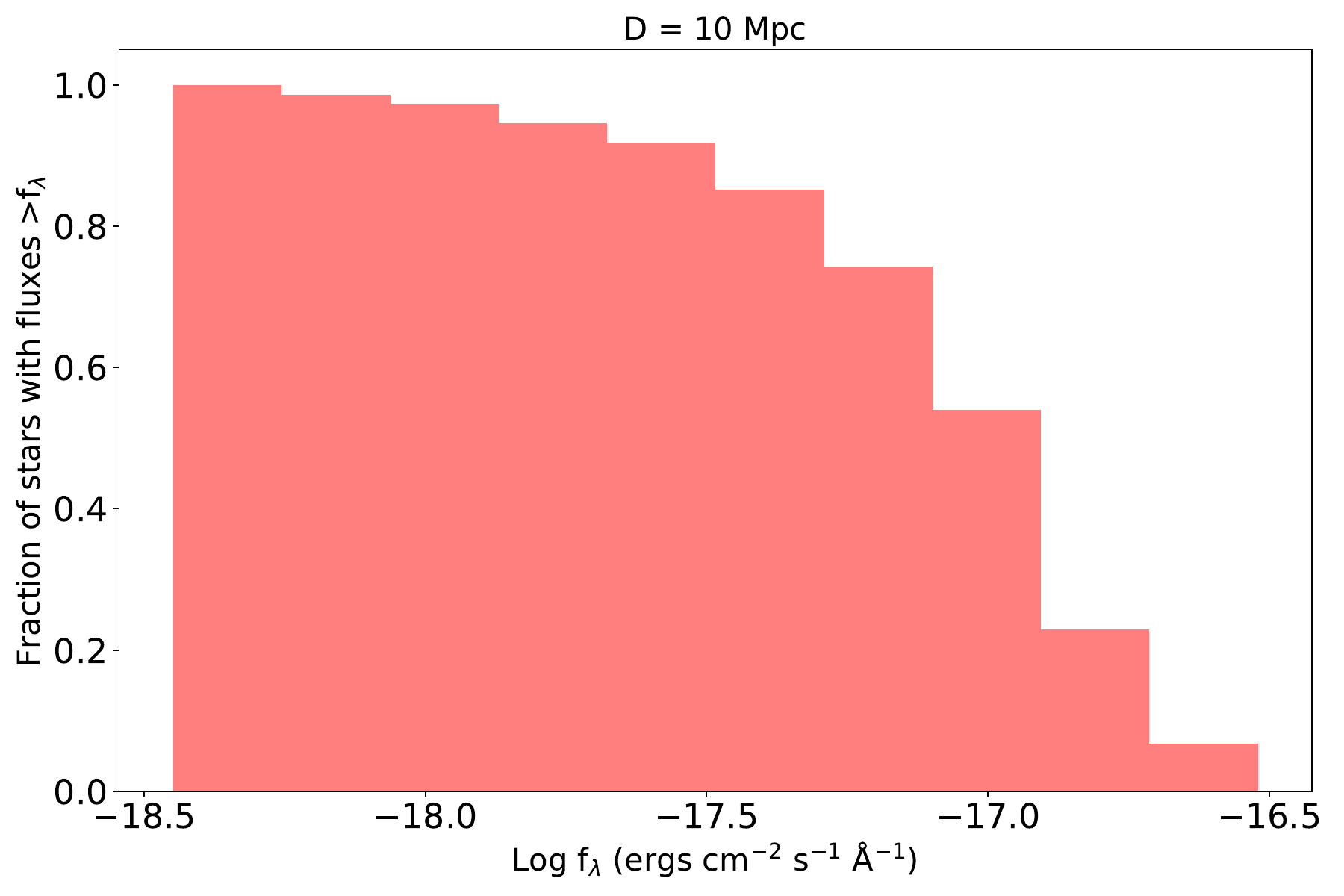}
\caption{\small Distribution of FUV (left) and NUV (right) fluxes of O and B stars in Sextans A (top), and Sextans A scaled at a distance of 10 Mpc (bottom). Fluxes are estimated based on the \citet{lorenzo2022} catalog of O and B stars in Sextans A (see Section \ref{sec_phys_param}).}
\label{fig4}
\end{center}
\end{figure*}

\begin{figure*}
\begin{center}
\includegraphics[width=8cm]{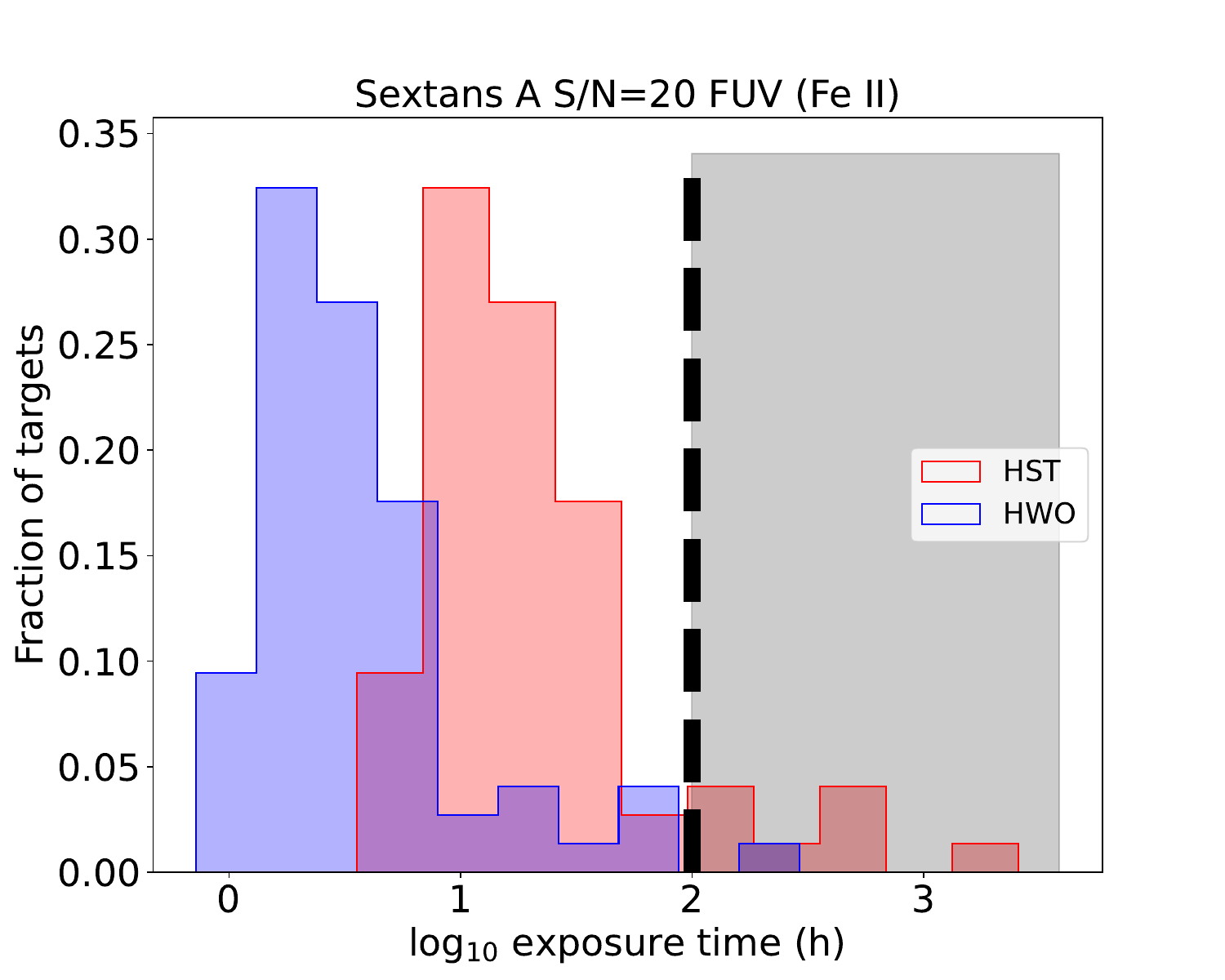}
\includegraphics[width=8cm]{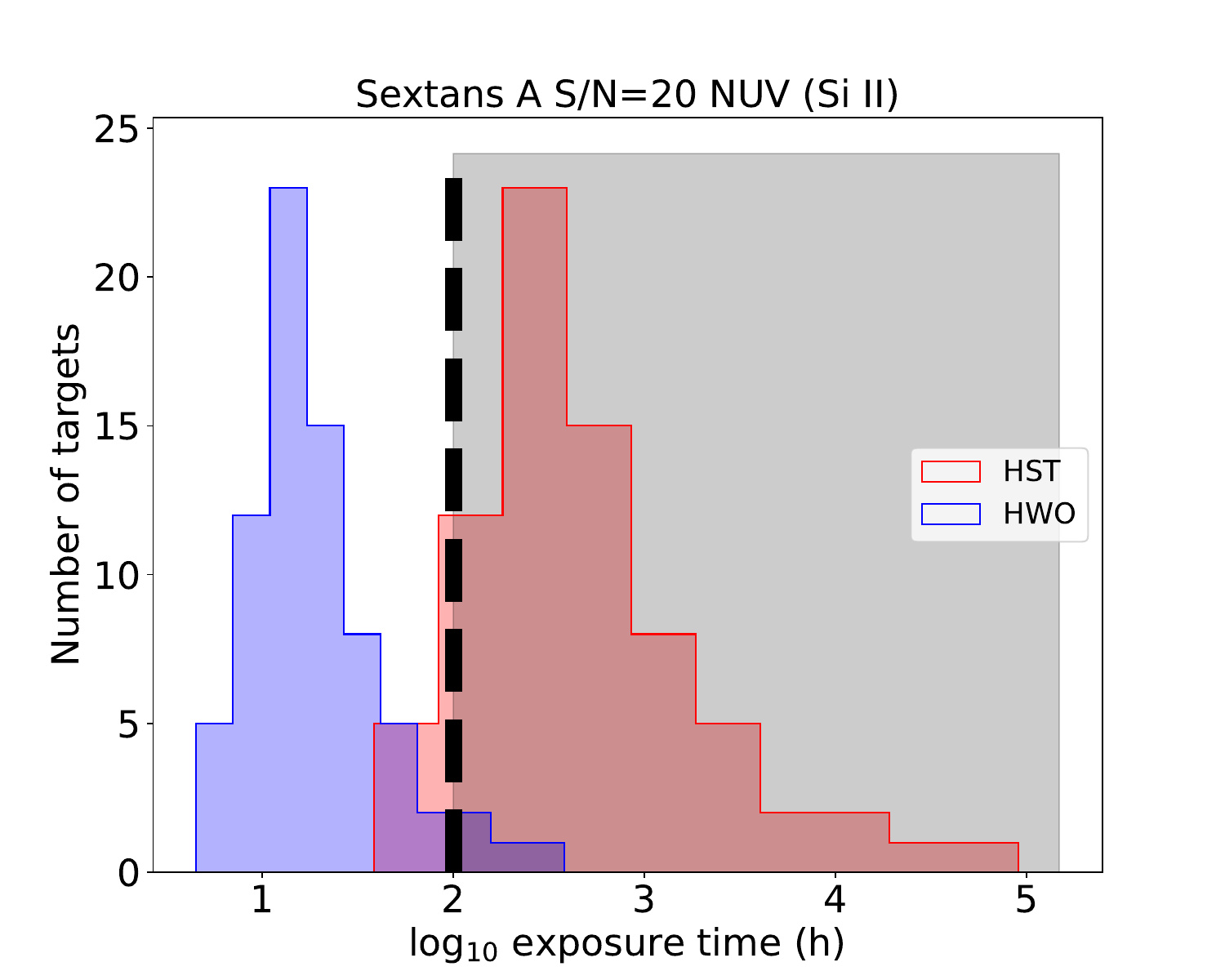}
\caption{\small Number of stars in the Lorenzo+2022 OB star catalogue (Sextans A, d = 1.3 Mpc) observable with HWO as a function of exposure time given the S/N and transition (Fe II 1144 \AA, Si II 1808 \AA). HWO will be able to observe most if not all the OB stars in Sextans A with R$\sim$50,000 and S/N $>$ 20 in less than 100h.
\label{fig5}}
\end{center}
\end{figure*}

\begin{figure*}
\begin{center}
\includegraphics[width=8cm]{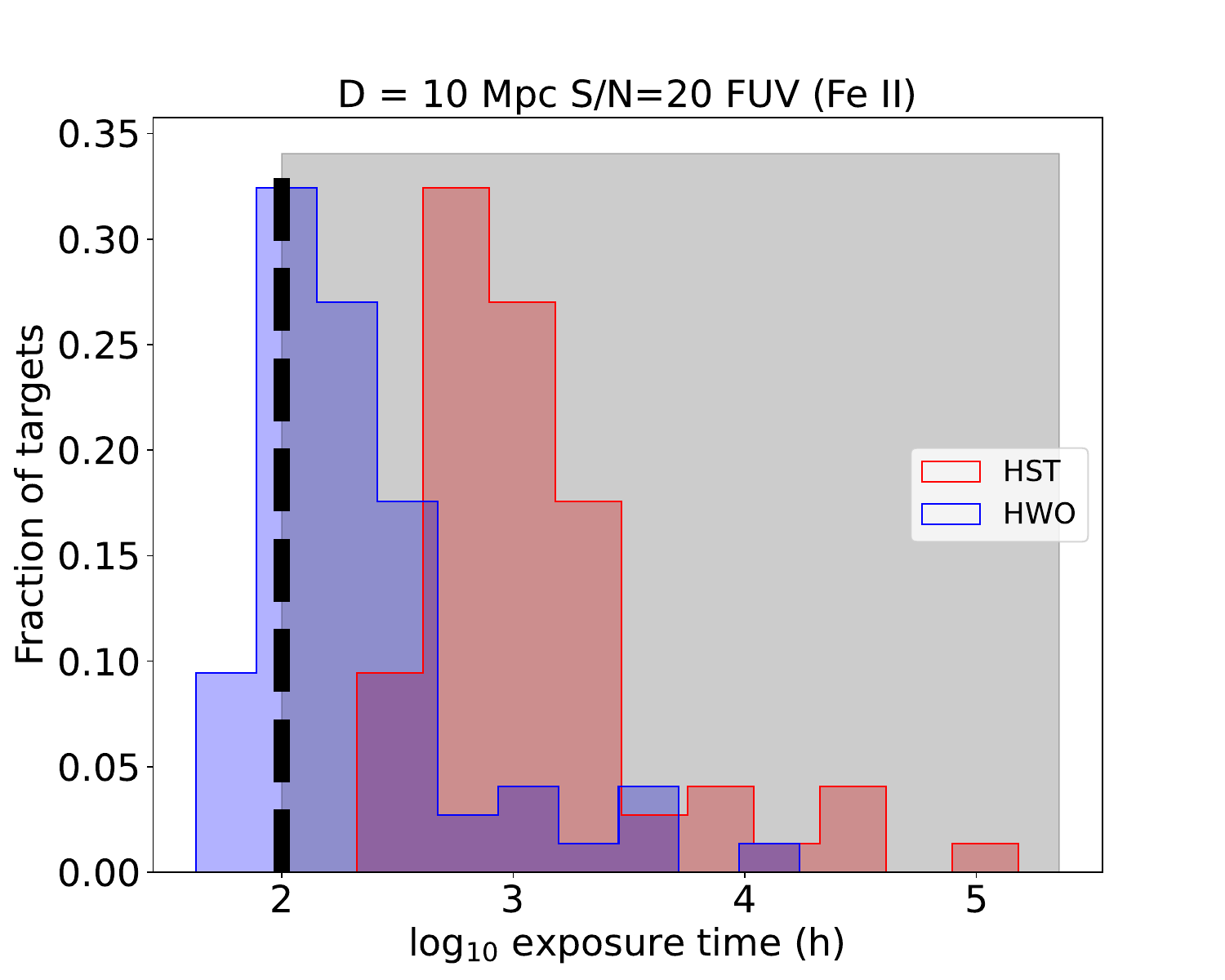}
\includegraphics[width=8cm]{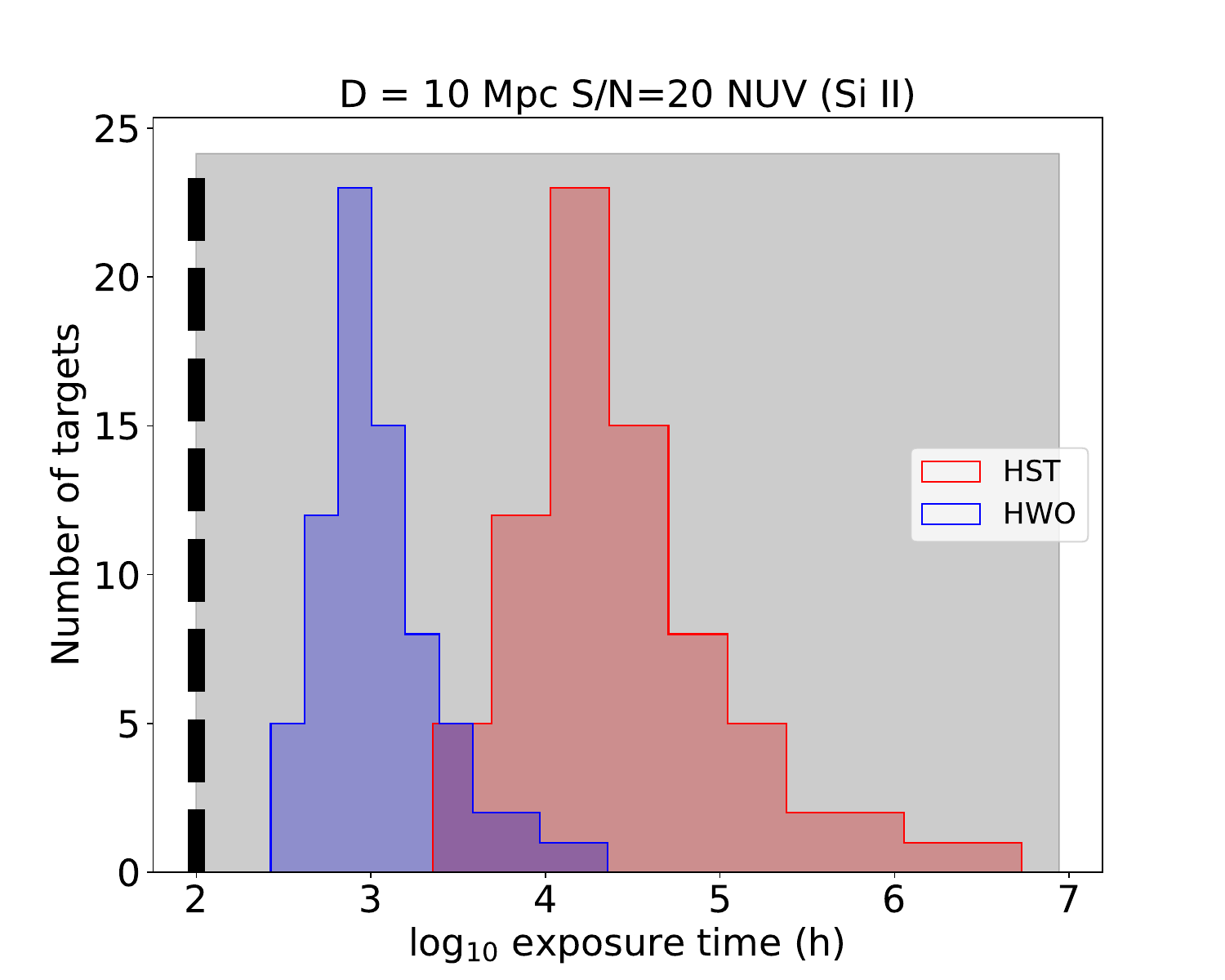}
\caption{\small Number of stars in the \citet{lorenzo2022} OB star catalogue scaled to a distance of 10 Mpc observable with HWO as a function of exposure time given the S/N and transition (Fe II 1144 \AA, Si II 1808 \AA). HWO will be able to observe a few dozens OB stars at 10 Mpc in the FUV with R~50,000 and S/N $>$ 20 in less than 100h. The NUV range appears to be beyond reach given the current sensitivity curves in the HWO ETC, but a factor of a few improvement would capture a few sightlines at 10 Mpc in the NUV.
\label{fig6}}
\end{center}
\end{figure*}

\indent The sightlines need to probe a range of gas densities, between log N(H) = 20 and 22 cm$^{-2}$ (as in state of the art studies in the LMC and SMC) and a range of metallicities, between 12 + log O/H = 7-8.6 (1-100\% solar). This requires the diversity of galaxies in the Local Volume (d $<$ 10 Mpc).\\
\indent We estimated the distribution of FUV (1250 \AA) and NUV (1800 \AA) fluxes in galaxies as a function of distance based on the \citet{lorenzo2022} catalog of O and B stars in Sextans A (d = 1.3 Mpc, Figure \ref{fig3}). First, the UV fluxes of Sextans A stars listed in \citet{lorenzo2022} were estimated using a \citet{castelli2004} model to which an LMC average dust extinction curve \citep{gordon2003} was applied with the E(B-V) specified in \citet{lorenzo2022}. The model spectrum was then normalized to the HST WFC3 F555W Vega magnitude listed in \citet{lorenzo2022}. Finally, all fluxes were scaled based on the distance considered. The resulting distributions of FUV and NUV fluxes in Sextans A and at 10 Mpc are shown in Figure \ref{fig4} \\
\indent Given the notional specifications of a LUMOS-like instrument (10x-100x more sensitive than Hubble depending on wavelength), the HWO ETC shows that S/N  = 12 could be achieved for $f_{\lambda}$ = 1$\times10^{-15}$ ergs cm$^{-2}$ s$^{-1}$ \AA$^{-1}$ in 1h for the FUV and 2.4h for the NUV. This implies that 2.8h and 8h are required to reach S/N = 20 in the FUV and NUV, respectively. Therefore, HWO would reach S/N $>$ 20 in $<$100h in the FUV and NUV for most OB stars in Sextans A  (d = 1.3 Mpc,  Figures \ref{fig3} \& \ref{fig5}). HWO should be able to reach the few brightest O and B stars in galaxies at d = 10 Mpc with S/N = 20 in the FUV ($f_{\lambda}$ $\sim$ a few times 10$^{-17}$ ergs cm$^{-2}$ s$^{-1}$ \AA$^{-1}$) and in $\sim$100h (Figure \ref{fig6}). The boundary of the Local Volume would be HWO's limit in terms of stellar UV spectroscopy, as Sextans A is for HST now.  The NUV range appears to be beyond reach for HWO at 10 Mpc given the current sensitivity curves in the HWO ETC (Figure \ref{fig6}), but a factor of a few improvement in NUV sensitivity may enable a few measurements for the brightest O and B stars at this distance. We note that the HWO ETC only includes sensitivity estimates below 2000 \AA, and we have not done any calculations for longer wavelengths as a result.\\
\indent Notionally, we would aim to observe 1-3 sight-lines per bin of 0.1 dex in column density (20-60 sightlines) for each galaxy. This would allow the derivation of the slope of the variations of depletions as function of ISM density with robust statistics. To probe the metallicity dependence of dust properties, 1-3 galaxies per bin of 0.1 dex in metallicity (20-60 galaxies) would be targeted. We note that there may not be enough galaxies in each metallicity bin to achieve this goal for all metallicities, particularly in the low metallicity range.\\

\subsection{Physical parameters for science objective \#2}

\indent This case requires very high resolution, high S/N UV spectroscopy, and due to the high dispersion, it is not expected that such an instrument have the sensitivity to reach massive stars beyond the Magellanic Clouds. Indeed, scaling the heavily reddened Milky Way star HD 37061 \citep[E(B-V) = 0.45; N(H) = 5.4$\times$10$^{21}$ cm$^{-2}$; N(CII) = 8.79$\times$10$^{17}$ cm$^{-2}$ N(CII*) = 4.65$\times$10$^{17}$ cm$^{-2}$; N(O) = 16.8$\times$10$^{17}$ cm$^{-2}$, see][]{sofia2004} to the distance of the SMC yields a NUV flux of 5$\times$10$^{-16}$ ergs cm$^{-2}$ s$^{-1}$ \AA$^{-1}$ and an FUV flux of a few times 10$^{-17}$ ergs cm$^{-2}$ s$^{-1}$ \AA$^{-1}$. The less reddened star HD36861 \citep[E(B-V) = 0.12; N(H) = 6.6$\times$10$^{20}$ cm$^{-2}$; N(C) = 0.83$\times$10$^{17}$ cm$^{-2}$; N(O) = 2.15$\times$10$^{17}$ cm$^{-2}$, see][]{sofia2004} scaled to the distance of the SMC has FUV and NUV fluxes of about 10$^{-13}$ ergs cm$^{-2}$ s$^{-1}$ \AA$^{-1}$. Assuming that the throughput of the high-resolution gratings on HWO is similar to that of the medium-resolution gratings, the limiting fluxes for S/N $=$ 100 and R = 100,000 would be 50$\times$ higher than the limiting fluxes for S/N = 20 at R = 50,000 (all in 100h), which corresponds to $f_{\lambda}$ $\sim$ a few times 10$^{-16}$ ergs cm$^{-2}$ s$^{-1}$ \AA$^{-1}$ in the FUV, and about $f_{\lambda}$ = 1$\times10^{-15}$ ergs cm$^{-2}$ s$^{-1}$ \AA$^{-1}$ in the NUV. Thus, HWO would not be able to observe SMC stars as reddened and with as much intervening carbon and oxygen columns as HD37061 (E(B-V) = 0.45) with R = 100,000 and S/N = 100. However, we would expect HWO to reach many O and B stars in the SMC with slightly lower extinction. HD36861 placed at the distance of the SMC would be easily reachable by HWO.\\
\indent For science objective \#2, large samples of O and B stars in the Milky Way and Magellanic Clouds (20-100\% solar metallicity) are required, probing gas column densities between log N(H) = 20 and 23 cm$^{-2}$.  We would aim to observe 1-3 sight-lines per bin of 0.1 dex in column density (20-60 sightlines) for each galaxy. \\

\section{Description of observations}

\subsection{Overall capabilities needed}

\indent For science objective \#1,  medium-resolution (R$\sim$50,000) multi-object spectroscopy (MOS) of massive stars in galaxies within $\sim$10 Mpc is needed over the full UV range (950 - 3150 \AA) and with S/N $>$ 20 in $<$100h for stars with fluxes as low as a few times 10$^{-17}$ ergs cm$^{-2}$ s$^{-1}$ \AA$^{-1}$ (estimate of the brighter end of OB star fluxes at $\sim$10 Mpc, see Figure \ref{fig4}).\\
\indent For science objective \#2,  the required capability is high-resolution (R $>$ 100,000), high S/N (S/N $>$ 100) multi-object or single object spectroscopy over the UV range 1150 - 3150 \AA\  toward massive stars in the Milky Way and Magellanic Clouds, with FUV and NUV fluxes as low as a few times 10$^{-16}$ ergs cm$^{-2}$ s$^{-1}$ \AA$^{-1}$. Sightlines with large column densities of carbon are heavily extincted and will have very low FUV fluxes compared to their NUV fluxes, due to the dust opacity being an order of magnitude higher in the FUV than in the NUV. With MOS, S/N $>$ 100 should be achieved in $<$ 100h.\\
\indent The observational requirements are summarized in Table \ref{tab:obs_requirements} for both science objectives.

\subsection{Aperture}

\indent The aperture should be at least 6m, to ensure that the brightest few stars at 10 Mpc can be observed with R $\sim$ 50,000 and S/N $>$ 20 in less than 100h in the FUV (those will have fluxes a few times 10$^{-17}$ ergs cm$^{-2}$ s$^{-1}$ \AA$^{-1}$). In the NUV, HWO may not be capable of acquiring these observations at 10 Mpc, but will reach at least 5 Mpc (Figure \ref{fig6}). This implies 10x-50x HST sensitivity in the FUV depending on wavelength, and $>$100x HST sensitivity in the NUV. A larger aperture would enable UV spectroscopic observations of larger samples of sightlines at 10 Mpc. Indeed, we expect a significant fraction of OB star fluxes at 10 Mpc to be only a few times 10$^{-18}$ ergs cm$^{-2}$ s$^{-1}$ \AA$^{-1}$ (Figure \ref{fig4}), making this kind of observations at 10 Mpc a stretch with the sensitivity in the HWO ETC, which assumes a 6m aperture (Figure \ref{fig6}). A larger aperture would also possibly enable NUV measurements (e.g., Zn, Fe) at 10 Mpc. 

\subsection{Spectral Resolution}

\indent R $\sim$ 50,000 is required to resolve ISM lines for science case \#1; R$>$100,000 is required for science case \#2 in order to robustly detect and analyze the weak O I and C II lines at 1355 and 2325 \AA. \\

\subsection{Spectral Range}

\indent For objective \#1, continuous UV coverage is needed from 950 \AA\ to get H$_2$ and P II lines, up to 3150 \AA\ to get Ly-$\alpha$, Mg II, Si II, Zn II, Fe II, Ni II, Cr II lines (and more, see full line list in \citet{ritchey2023}. In addition to the FUV range, which includes many transitions for hydrogen and most dust constituents, the NUV range is critical to obtain the key Si II (1808 \AA) and Zn II (2026, 2062 \AA). Si is a main dust constituent, and Zn is key to deriving depletion corrections based on abundance ratios to be used to track the chemical enrichment of the universe through DLAs.\\
\indent For objective \#2, the required UV range is 1150-3150 \AA\ to get O I (1355 \AA), and C II (2325 \AA). Ly-$\alpha$ can be obtained with medium-resolution, but measuring abundances in the Milky Way using other transitions (e.g., S II $\lambda$$\lambda$$\lambda$ 1250, 1253, 1259) could significantly benefit from the highest resolution.

\subsection {Other}

\indent MOS is preferred over IFU ($>$ 100 apertures for large samples of stars). Objective \#2 requires the ability to reach high S/N ($>$100).  This is only possible with a stable and repeatable P-flat. HST/COS micro-channel plates (MCPs) cannot achieve this level of S/N due to fixed pattern noise. This may require fixed-pixel detectors.  Note that the high-resolution gratings (E140H, E230H) on HST/STIS have produced almost 200 refereed papers. It would be hard to imagine HWO missing this opportunity.

\subsection{Need for HWO}

\indent The state-of-the-art, also described in previous sections, was achieved by the COS and STIS spectrographs on Hubble. Those instruments have enabled measurements of neutral gas-phase abundances and dust depletions for the main constituents of dust except C and O (Fe, Si, Mg, Cr, Ni etc.), in the Milky Way and Magellanic Clouds (20-100\%), across a broad ISM density range (two orders of magnitude). Those measurements utilized medium-resolution single object spectroscopy toward background O and early B stars. Thanks to these measurements, variations in the dust abundance and properties are relatively well understood in those specific environments. However, the contribution of C and O to the dust content has not been measured outside the Milky Way. And galaxies beyond the Milky Way and Magellanic Clouds, in particular very low metallicity environments and starbursts that are analogs to the high redshift universe, are beyond the reach of Hubble. \\
\indent Obtaining neutral-gas phase abundance and dust depletions for the full suite of dust constituents, and over a large enough number of sightlines that the dependance of the dust abundance and properties on local environment, in particular ISM density, can be constrained, will require the sensitivity, spectral resolution, wavelength coverage, and multiplexing capabilities of HWO. In particular, solving the problem of the dust composition and carbon/oxygen content of dust at low metallicity requires high-resolution (R$>$100,000) gratings in the FUV and NUV and the ability to reach S/N $>$ 100.

\begin{table*}[!ht]
\begin{center}
  \begin{tabular}{p{0.19 \linewidth} | p{0.19\linewidth} | p{0.19\linewidth} |  p{0.19\linewidth}  | p{0.19\linewidth}  }
Observation requirement& State of the Art  & Incremental Progress (Enhancing) &Substantial progress (enabling)  & Major progress (breakthrough)    \\
\hline
&&&& \\
Type (imaging, spectroscopy, etc.) & Single object Spectroscopy  \newline \newline  HST/COS/FUV \newline \newline HST/COS/NUV \newline \newline HST/STIS/FUV \newline \newline HST/STIS/NUV & Single object spectroscopy & Multi-object spectroscopy & Multi-object spectroscopy \\
\hline
&&&& \\
Wavelength range &  950-3150  \AA, but very low throughput below 1100 \AA\ and  low throughput above 1750  \AA\ \newline \newline COS/NUV has very narrow bandpasses requiring multiple settings & 950 - 3150 \AA\ & 950 - 3150 \AA\  with relatively uniform throughput (no cliff at 1100 \AA) &  950 - 3150 \AA\  with relatively uniform throughput (no cliff at 1100 \AA) \\
\hline
&&&& \\
Number of targets & see Table 1 & see Table 1 & see Table 1 & see Table 1\\
&&&& \\
\hline
&&&& \\ 
Magnitude of target in chosen bandpass &  COS/FUV: limiting flux 10$^{-15}$ ergs cm$^{-2}$ s$^{-1}$ \AA$^{-1}$ \newline \newline  COS/NUV: limiting flux 5$\times$10$^{-15}$ ergs cm$^{-2}$ s$^{-1}$ \AA$^{-1}$ \newline \newline STIS/FUV/R=100,000 limiting flux 5$\times$10$^{-14}$ ergs cm$^{-2}$ s$^{-1}$ \AA$^{-1}$ \newline \newline STIS/NUV/ R=100,000  limiting flux 1.5$\times$10$^{-13}$ ergs cm$^{-2}$ s$^{-1}$ \AA$^{-1}$ \newline \newline Limiting fluxes for S/N $>$ 20 in $<$10h   &   At R$\sim$20,000: \newline \newline  FUV: 5$\times$10$^{-16}$ ergs cm$^{-2}$ s$^{-1}$ \AA$^{-1}$ \newline \newline NUV: 2.5$\times$10$^{-15}$ ergs cm$^{-2}$ s$^{-1}$ \AA$^{-1}$ \newline \newline  At R$\sim$100,000 \newline \newline FUV: 2.5$\times$10$^{-14}$ ergs cm$^{-2}$ s$^{-1}$ \AA$^{-1}$ \newline \newline NUV: 0.8x10$^{-13}$ ergs cm$^{-2}$ s$^{-1}$ \AA$^{-1}$ \newline \newline (Factor 2 improvement in sensitivity) \newline \newline Limiting fluxes for S/N $>$ 20 in $<$10h   &  At R$\sim$50,000 \newline \newline FUV: 2$\times$10$^{-16}$ ergs cm$^{-2}$ s$^{-1}$ \AA$^{-1}$ \newline \newline  NUV: 3$\times$10$^{-16}$ ergs cm$^{-2}$ s$^{-1}$ \AA$^{-1}$ \newline \newline   At R$\sim$100,000 \newline \newline FUV: 5$\times$10$^{-15}$ ergs cm$^{-2}$ s$^{-1}$ \AA$^{-1}$ \newline \newline NUV: 5$\times$10$^{-15}$ ergs cm$^{-2}$ s$^{-1}$ \AA$^{-1}$ \newline \newline Limiting fluxes for S/N $>$ 20 in $<$10h &  At R$\sim$50,000 \newline \newline  FUV: 3$\times$10$^{-17}$ ergs cm$^{-2}$ s$^{-1}$ \AA$^{-1}$ \newline \newline   NUV: 1.6$\times$10$^{-17}$ ergs cm$^{-2}$ s$^{-1}$ \AA$^{-1}$ \newline \newline  At R$\sim$100,000 \newline \newline FUV: 5$\times$10$^{-16}$ ergs cm$^{-2}$ s$^{-1}$ \AA$^{-1}$ \newline \newline NUV: 5$\times$10$^{-16}$ ergs cm$^{-2}$ s$^{-1}$ \AA$^{-1}$ \newline \newline  Limiting fluxes for S/N $>$ 20 in $<$10h\\
\hline
&&&& \\
Spectral resolution R &  COS/FUV+NUV: R=20,000 \newline \newline STIS/FUV+NUV: R = 30,000-100,000 &  R$\sim$20,000 \newline \newline R$\sim$100,000 &  R$\sim$50,000 \newline \newline R$\sim$100,000 &  R$\sim$50,000 \newline \newline R$\sim$100,000\\
\end{tabular}
\end{center}
\caption{Observational requirements for the state-of-the-art, as well as incremental, substantial, and major progress scenarios.}
\label{tab:obs_requirements}
\end{table*}

{\bf Acknowledgements.} Based on observations obtained with the NASA/ESA Hubble Space Telescope, retrieved from the Mikulski Archive for Space Telescopes (MAST) at the Space Telescope Science Institute (STScI). STScI is operated by the Association of Universities for Research in Astronomy, Inc. under NASA contract NAS 5-26555.

\bibliography{biblio_all}{}

\end{document}